\documentclass[slac_one]{revtex4}
\usepackage{graphicx}
\usepackage{fancyhdr}
\usepackage{bm}        
\usepackage{amssymb}   
\usepackage{amsmath}   
\usepackage{xspace}
\def\ppbar{$p\overline{p} $}            
\def\D0{D\O}                            
\newcommand{\Ptg}{p_{T}^{\gamma}}
\newcommand{\rJLIP}{\ensuremath{P_{b\text{-jet}}}\xspace}
\newcommand{\GeV}{\ensuremath{\text{GeV}}\xspace}

\newcommand{\gb}{\ensuremath{\gamma+{b}}\xspace}
\newcommand{\gc}{\ensuremath{\gamma+{c}}\xspace}
\begin{document}

\title{Measurements of Photon Plus Heavy Flavor Jet Cross Sections} 

%

\author{D. Duggan (for the \D0 Collaboration)}
\affiliation{Florida State University, Tallahasee, FL 32306, USA}
%

\begin{abstract}
Measurements of differential photon $+ c$ jet and photon $+ b$ jet production
cross sections are presented using approximately 1 $fb^{-1}$ of data collected by 
the \D0 detector at the Tevatron \ppbar~collider at a center-of-mass energy of
1.96 TeV. Isolated photons are selected in the rapidity range $|y^\gamma| < 1.0$
and jets selected with rapidities $|y^\text{jet}| < 0.8$. The measurements 
are compared to next-to-leading order theoretical predictions.
\end{abstract}

\maketitle

\thispagestyle{fancy}

\section{INTRODUCTION} 
We present first preliminary measurements of differential $\gamma + $ c
jet and $\gamma + $ b jet cross sections using $\sim1~fb^{-1}$ of data. These measurements
were performed at the Fermilab Tevatron \ppbar~collider at $\sqrt{s}$ = 1.96 TeV 
collected using the \D0 detector. These measurements are differential with respect to the transverse
momentum of the leading photon ($\Ptg$), the rapidity of the leading jet ($y^\text{jet}$) 
and of the leading photon ($y^{\gamma}$). The results are presented in five $\Ptg$ bins and
in two regions of photon--\,jet rapidities: $y^\gamma \cdot y^\text{jet} > 0$ (region 1)
and $y^\text{jet} \cdot y^{\gamma} < 0$ (region 2).
\section{DATA SELECTION}
Events are recorded using the \D0 detector~\cite{D0_det} and must contain 
at least one photon candidate and at least one heavy flavor jet
candidate. Both the chosen photon and jet candidates are required to be leading with
respect to their $p_{T}$. The photon candidate $p_{T}$ must be greater than 30 GeV,
be located within the central calorimeter and have $|y^{\gamma}| < 1.0$. The leading
jet must have $p_{T} > 15$ GeV and $|y^\text{jet}| < 0.8$. Additionally,
events must also contain a primary vertex within 50~cm of the detector's
center along the beam pipe. This vertex is required to have at least three associated tracks,
one with $p_{T} > 1.0$ GeV and a second with $p_{T} > 0.5$ GeV. To suppress
events containing cosmic muons and W bosons, the missing transverse energy
($E_{T}^{Miss}$) must pass the condition $E_{T}^{Miss} < 0.7 \cdot \Ptg$.

Events are triggered using electromagnetic (EM) clusters consistent with photon shower profiles.
The final photon energy is taken as the energy of the cluster within a 
cone of radius $R = 0.2$. An isolation requirement is imposed on the candidate photon 
such that the total energy in a cone of $R = 0.4$, excluding $E_{EM}$, must be less than 7\%
of $E_{EM}$. The cluster must also have a probability less than 0.1\%
to be spatially matched to any track in the event. The cluster's transverse profile
in the third layer of the EM calorimeter must be consistent with that of a photon.
The primary vertex is further constrained to be within 10~cm of the most
probable origin of the photon candidate along the beam pipe. To further suppress 
backgrounds coming from jets, an artificial neural network (ANN)~\cite{gamjet_PLB}, 
which uses additional single variable discriminants, was employed with the requirement 
of ANN $> 0.7$. 

The energy of the leading jet is determined using \D0's RunII jet reconstruction
algorithm~\cite{c:Run2Cone} with a cone of radius $R = 0.5$. 
Heavy flavor jets are identified using their long lifetimes by reconstructing 
displaced secondary vertices from the jet's tracks.
To best enhance the fraction of jets originating from heavy flavor quarks, an artificial
neural network (bANN) was implemented. The bANN~\cite{c:bNN} uses variables characteristic
of heavy flavor jets as inputs, trained such that the output of $b$ jets 
tends toward one and that of light (udsg) jets tends to zero. For this selection,
we require the bANN output $> 0.85$.

\section{PURITY ESTIMATES}
\begin{figure}
                                                                                
\includegraphics[width=0.40\linewidth]{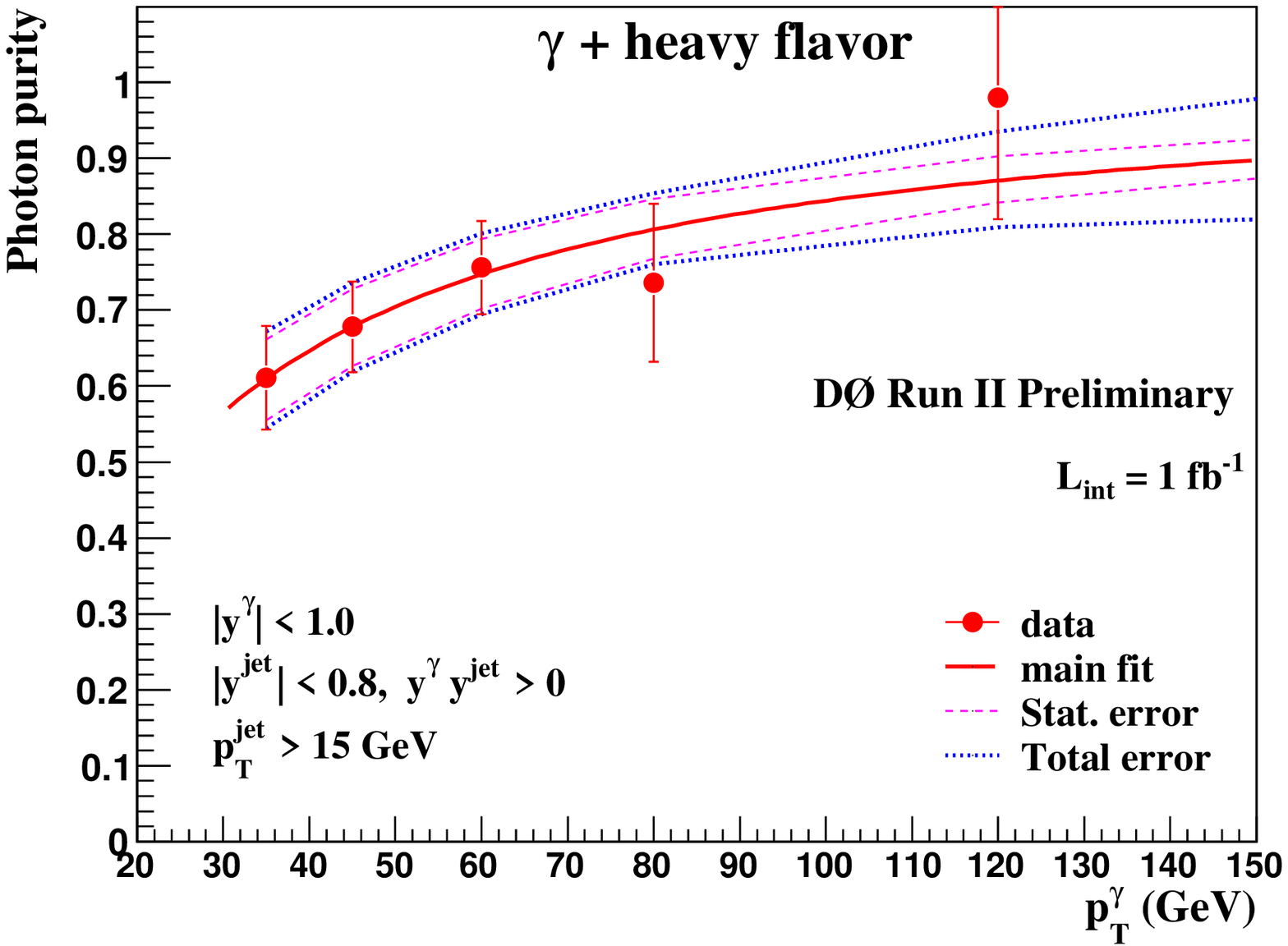}
\includegraphics[width=0.40\linewidth]{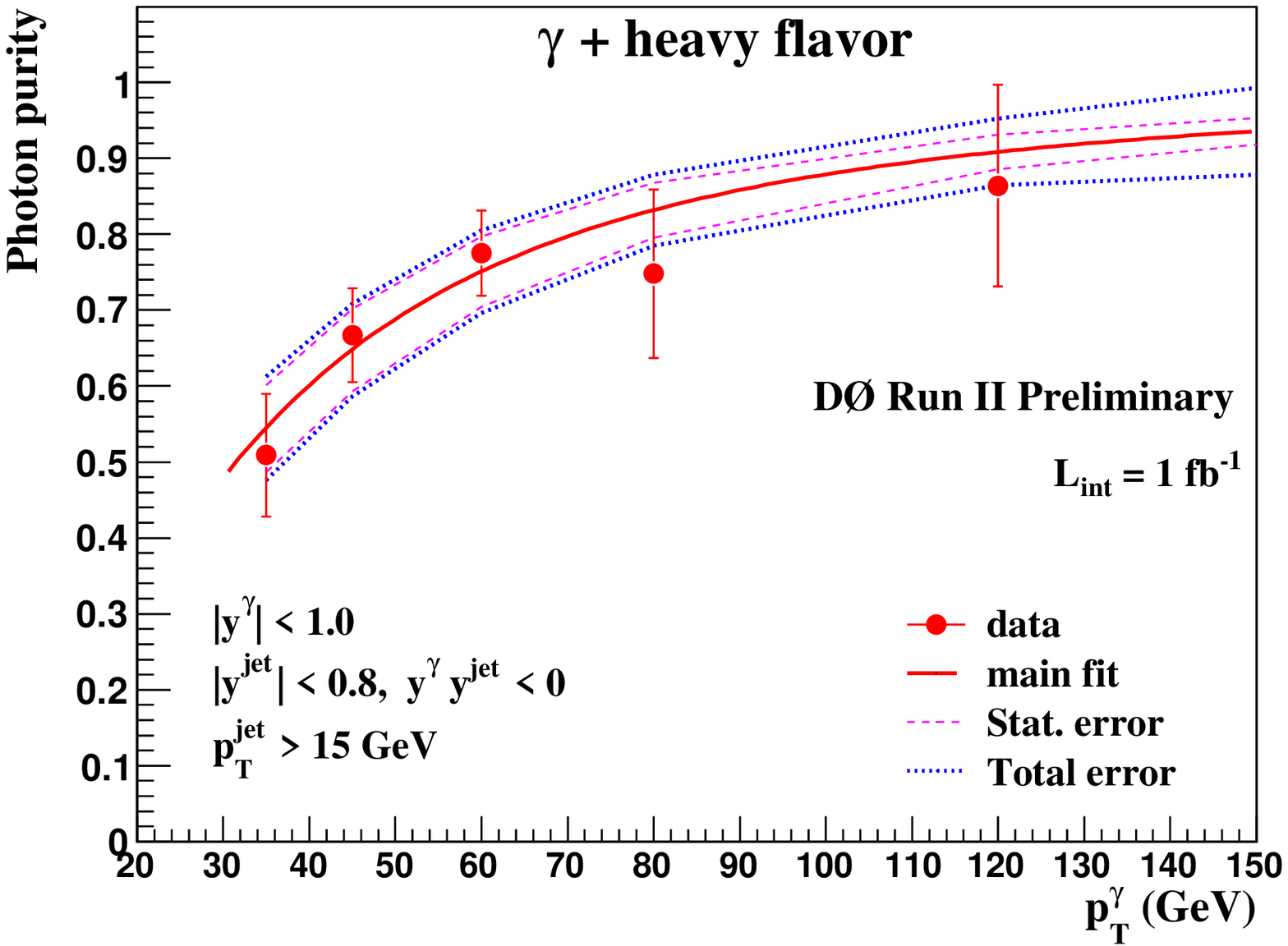}
                                                                                
\caption{The photon purity in the selected $\gamma$ + heavy flavor jet
events as a function of $\Ptg$ for both rapidity regions 1 (left) 
and 2 (right). The result of the
$\left(1 - \exp(a + b p_T^\gamma)\right)$ functional fit is shown by
the full lines, together with the statistical uncertainty in the
default fit (dashed lines) and the total uncertainty (dotted lines).}
\label{fig:pur_fit}
\end{figure}

To estimate the fraction
of photons in the final data sample, the photon ANN output in data is fit with
a template from signal photon Monte Carlo (MC) {\sc pythia}~\cite{PYT} simulation
and a background dijet MC template.
The fit is performed in each $\Ptg$ bin and in each photon-jet rapidity region
separately. The fit results are shown in Fig.~\ref{fig:pur_fit}.

\begin{figure}
\centering
\includegraphics[width=0.40\linewidth,clip= true,bb= 0 400 268 596]{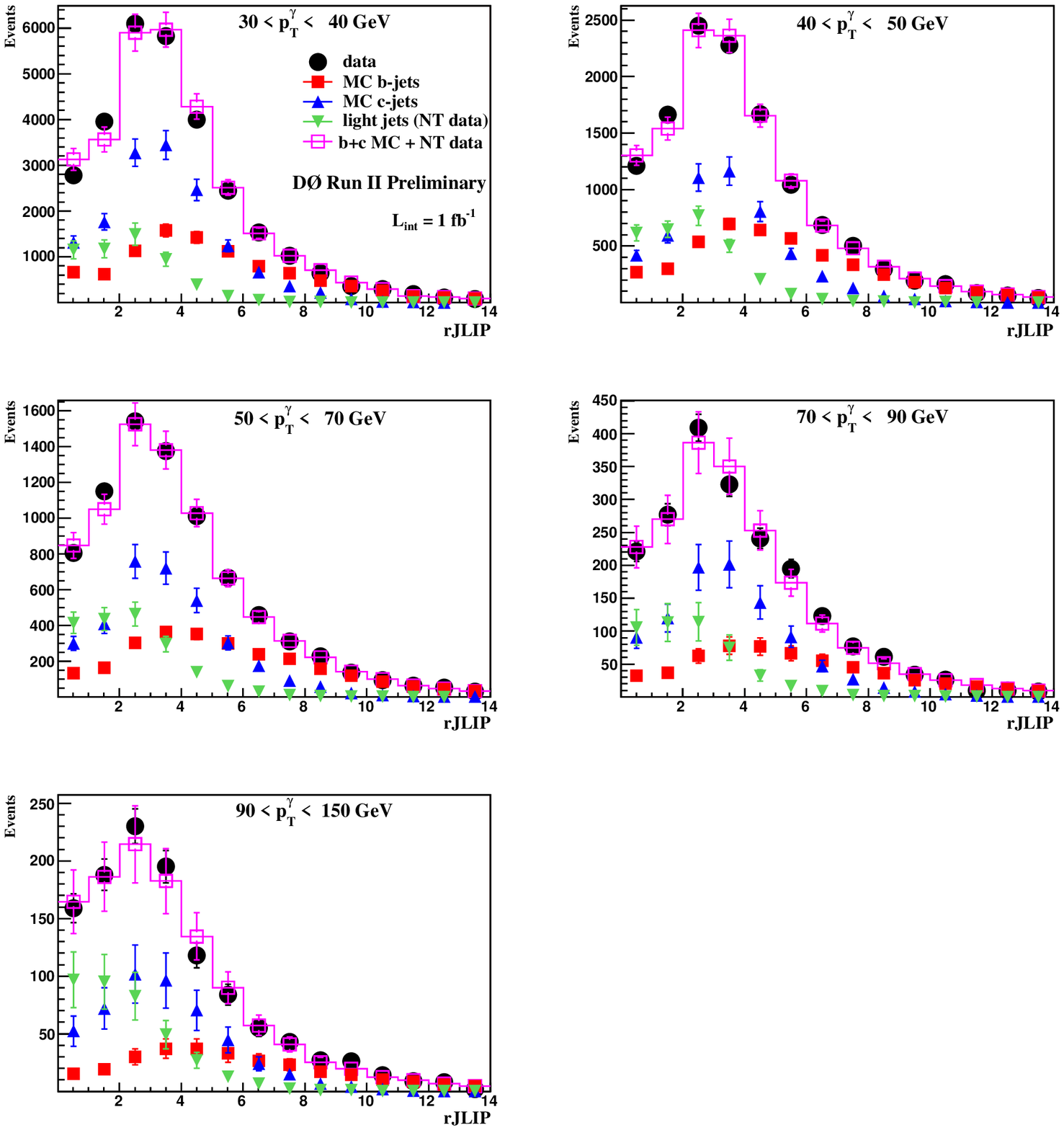}
\caption{Distribution of the number of events in data for \rJLIP after
  all selection criteria in the $\Ptg$ interval 30 to 40~\GeV.
  The $b$, $c$ and light jet template distributions are also shown
  normalized to their fitted fraction.  The error bars on the data
  points show the combined statistical uncertainty from the data
  statistics and the flavor fraction fit.}
\label{fig:cbjet_test}
\end{figure}

To estimate the fraction of $c$ and $b$ jets in the final data sample,
a different discriminant was used. It is defined as
$\rJLIP=-\ln\prod_{i}{P_{\rm track}^{i}}$, where $P_{\rm track}^{i}$ is the probability
of a track in the jet cone to originate from the primary vertex,
omitting the track with the lowest probability.
\rJLIP has large values for $b$ jets and values closer to zero for light
jets. A fit to the data \rJLIP distribution is performed using MC
templates for $b$ and $c$ jets, and the light jet template is taken from
a light jet enriched data sample. The fit done in each $\Ptg$ bin, and
to verify the quality of the fit, a combined flavor template is compared
to the data as shown in Fig.~\ref{fig:cbjet_test}.
\section{CROSS SECTION RESULTS}
\begin{figure}
\centering
\includegraphics[width=0.35\linewidth]{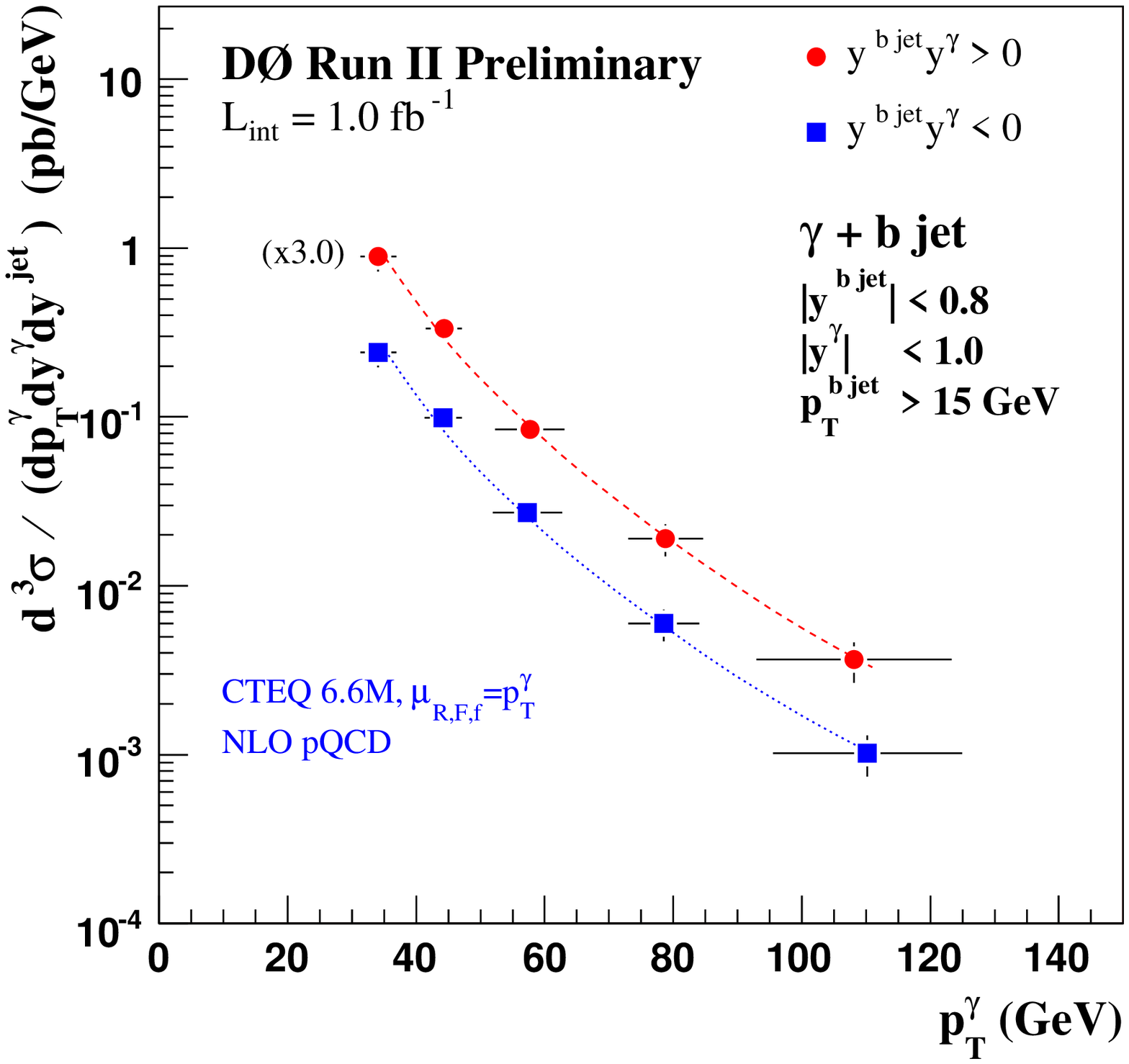}
\includegraphics[width=0.35\linewidth]{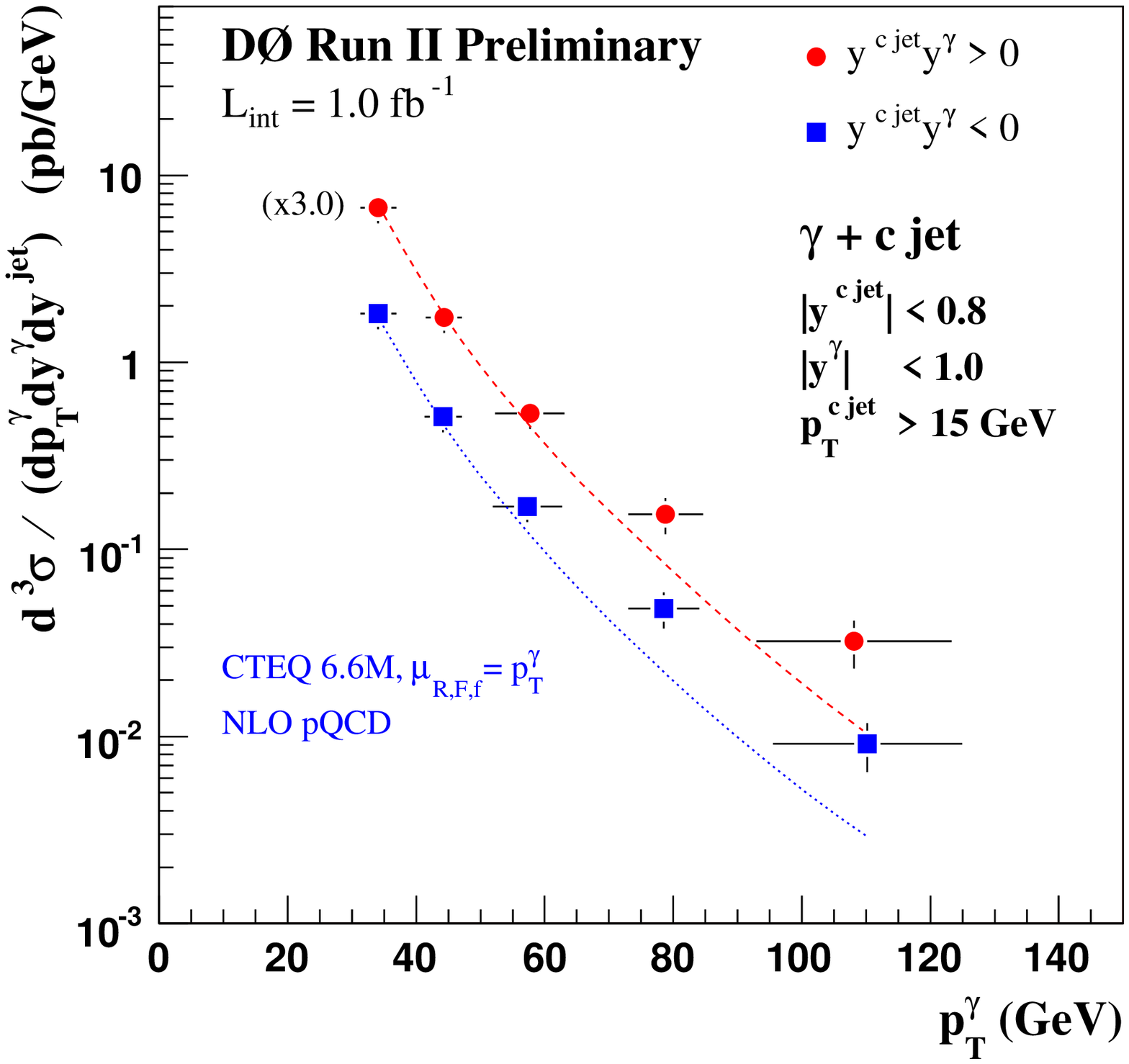}
\caption{ The \gb~(left) and \gc~(right) cross sections as a
  function of $\Ptg$ in regions~1
  and 2.  The uncertainties on the points in data are the full
  uncertainties. The NLO theoretical predictions using
  the {\sc cteq}6.6M PDF set are shown by the dotted lines.}
\label{fig:xsectbc1plot}
\end{figure}

The differential cross sections are presented in five $\Ptg$ bins and two regions
of $y^\gamma \cdot y^\text{jet}$, corrected for $\Ptg$
smearing effects due to the finite resolution of the calorimeter using
the unfolding method described in~\cite{D0_unsmearing}.  The measured
cross sections are shown in Fig.~\ref{fig:xsectbc1plot} for \gb
and \gc processes. 
Statistical uncertainties on the results vary from 0.2\,\% in the
lowest $\Ptg$ bin to 8--9\,\% in the highest bin, while systematic
uncertainties vary between 15--25\,\%.
The theoretical predictions from next-to-leading (NLO) order calculations
are presented in Fig.~\ref{fig:xsectbc1plot} with the renormalization,
factorization and fragmentation scales $\mu_{R}, \mu_{F}$, and $\mu_f$
set to $\Ptg$.  These predictions are preliminary~\cite{Tzvet} and are
based on techniques to calculate the analytic cross sections
published in Ref.~\cite{Harris}.

\begin{figure}
\includegraphics[width=0.35\linewidth]{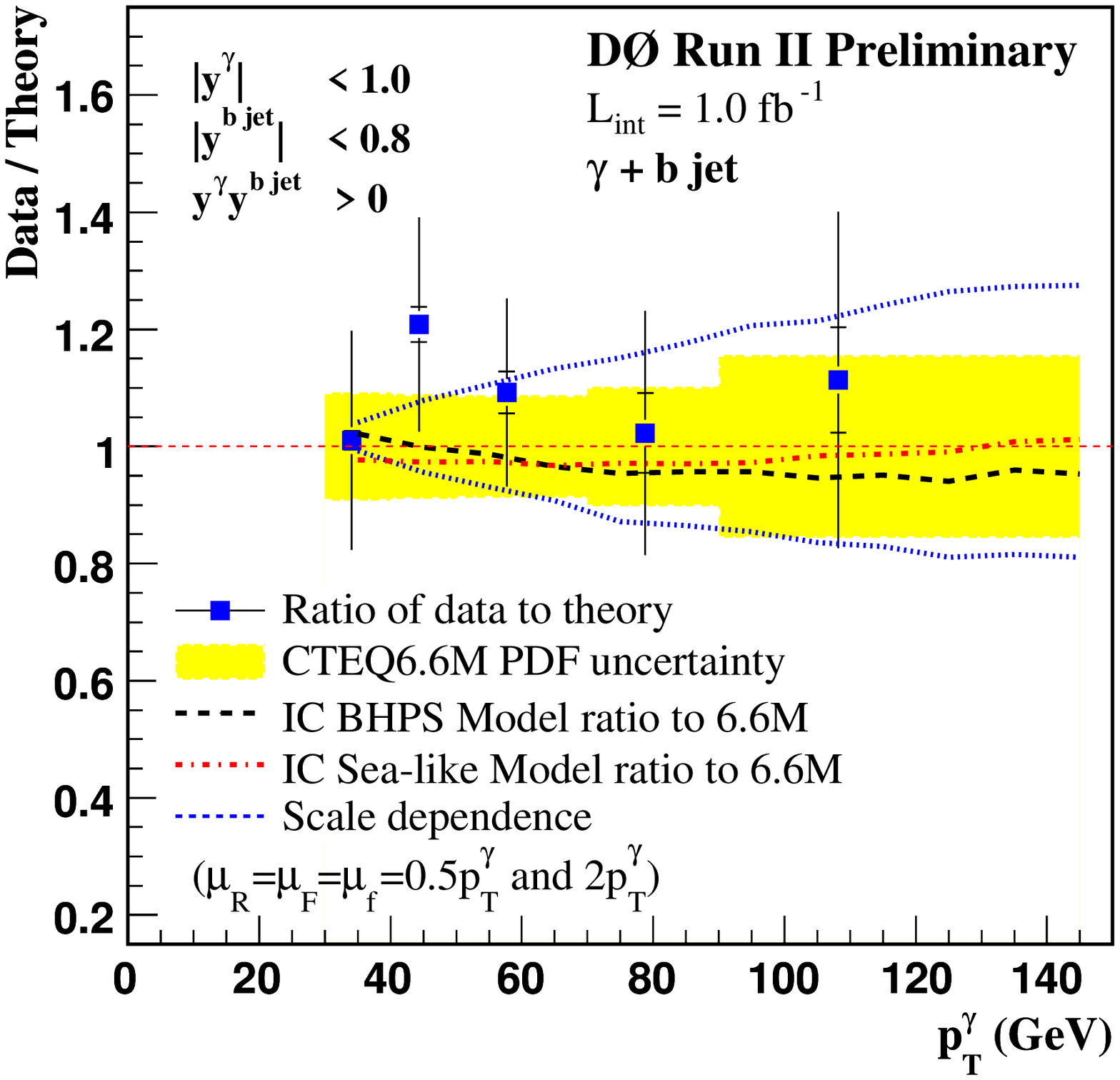}
\includegraphics[width=0.35\linewidth]{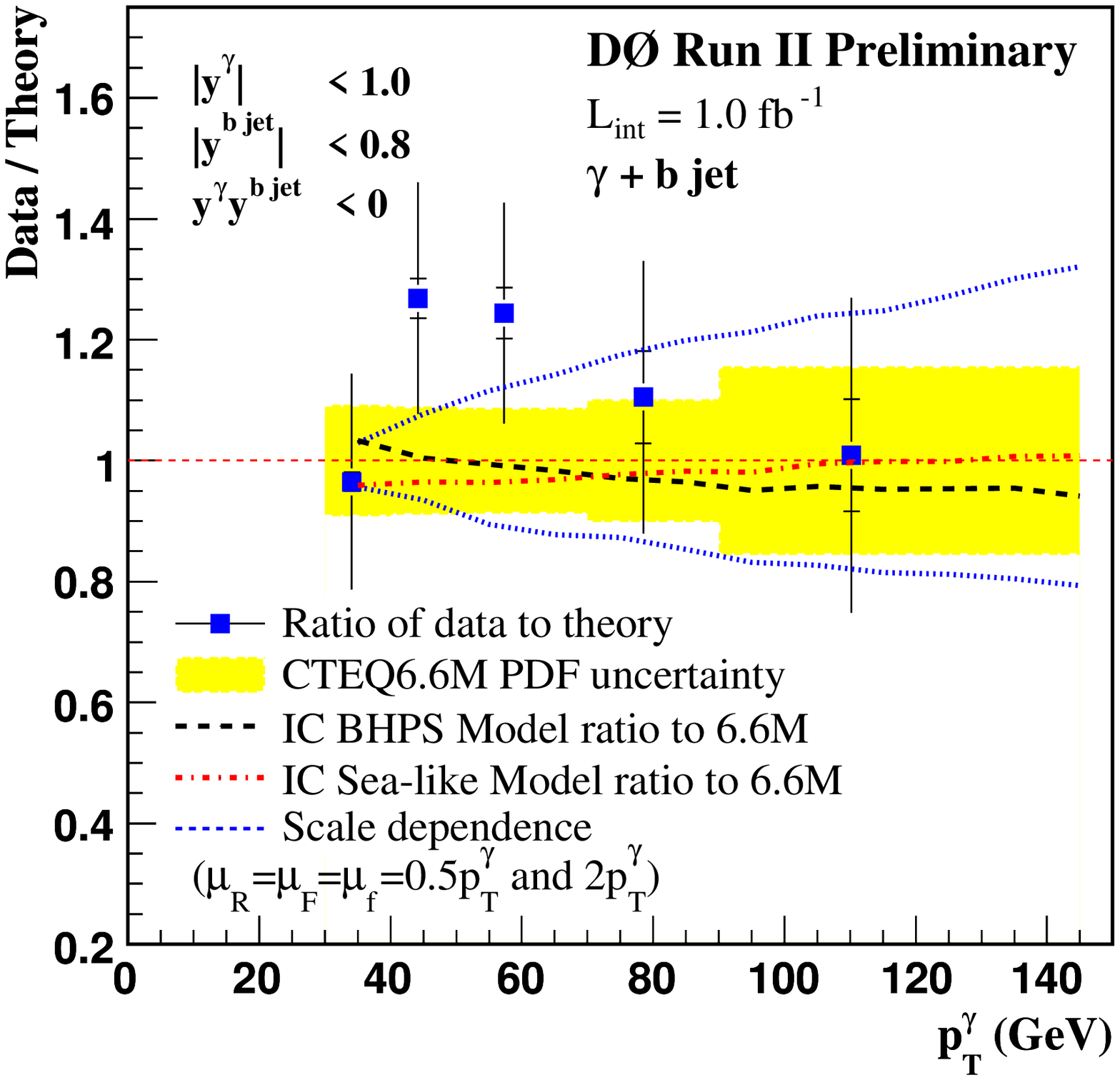}
                                                                                
\includegraphics[width=0.35\linewidth]{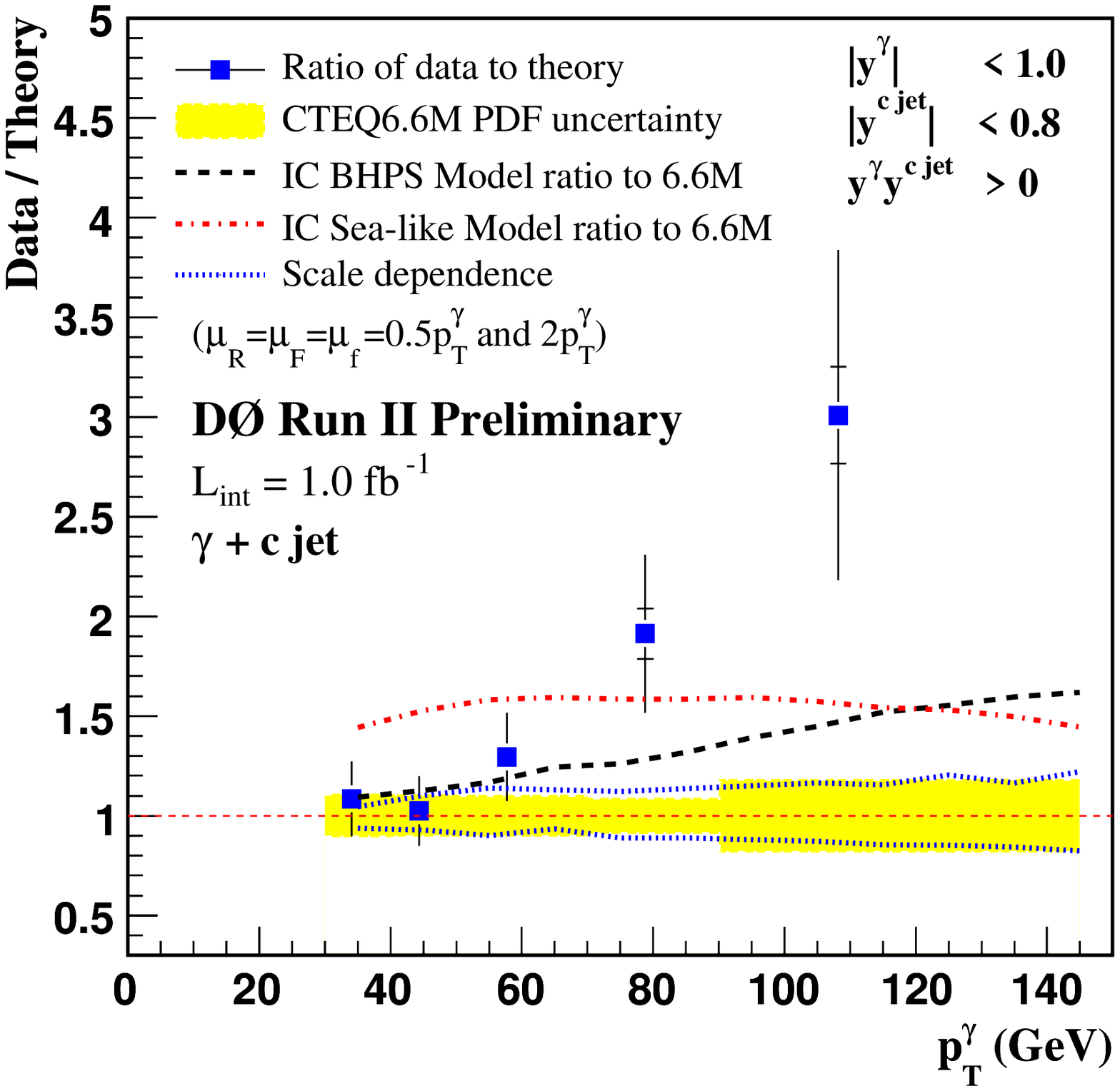}
\includegraphics[width=0.35\linewidth]{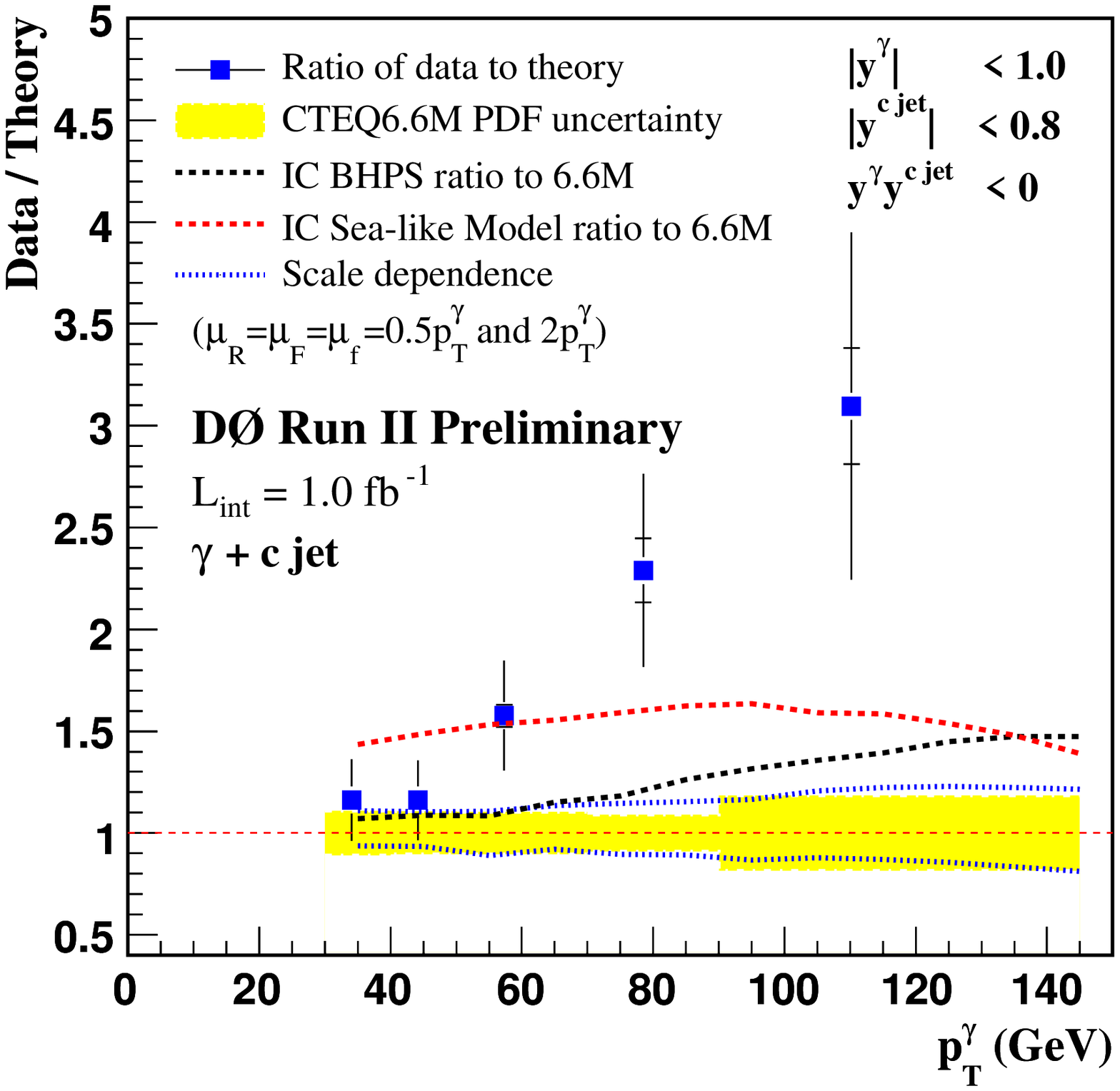}
\caption{ The \gb cross section ratio of data to theory as a
  function of $\Ptg$ in region~1 (top left) and region~2 (top right)
  and the \gc cross sections in region~1 (bottom left) and region~2
  (bottom right).  The uncertainties on the data points include both
  statistical (inner line) and full uncertainties (entire error bar)
  Also shown are the theoretical scale uncertainties and the {\sc
    cteq}6.6M PDF uncertainties. The ratio of the central {\sc
    cteq}6.6M prediction to two \emph{intrinsic charm} models is shown
  as well.}
\label{fig:xsectb1ratio}
\end{figure}

The ratios of the measured and predicted cross sections
for both \gb and \gc cross sections in the two
rapidity regions are shown in Fig.~\ref{fig:xsectb1ratio}.  The
uncertainty due to the scale choice is estimated by
simultaneously varying all three scales by a factor of two,
$\mu_{R,F,f}=0.5 p_T^\gamma$, and $\mu_{R,F,f}=2 p_T^\gamma$.  The
theoretical predictions utilize the {\sc cteq}6.6M PDF set with
uncertainties calculated according to the
prescription in \cite{CTEQ}. 

\begin{acknowledgments}
The author is grateful to Tzvetalina Stavreva and Jeff Owens and
thanks the staffs at Fermilab and collaborating institutions.
\end{acknowledgments}

\end{document}